\documentclass[sigconf,nonacm]{acmart}

\AtBeginDocument{}

\usepackage{booktabs}
\usepackage{multirow}
\usepackage{array}
\usepackage{amsmath}
\usepackage{enumitem}
\usepackage{xspace}
\usepackage{balance}
\usepackage{graphicx}
\usepackage{algorithm}
\usepackage{algorithmic}
\usepackage{microtype}
\usepackage{tabularx}
\usepackage{orcidlink}

\setcopyright{none}
\renewcommand\footnotetextcopyrightpermission[1]{}

\newcommand{\system}{EduGuard\xspace}
\newcommand{\dataset}{BILearn-CS\xspace}
\title[EduGuard]{EduGuard: A Safe RAG-Based LLM Tutor for Programming Education}

\author{S M Asif Hossain}
\orcid{0009-0002-4749-109X}

\author{Ruksat Khan Shayoni}
\orcid{0009-0002-8737-1789}

\author{M. F. Mridha}
\orcid{0000-0001-5738-1631}

\author{Jungpil Shin}
\orcid{0000-0002-7476-2468}

\renewcommand{\shortauthors}{Hossain et al.}

\makeatletter
\def\@mkauthors{%
  \global\setbox\mktitle@bx=\vbox{%
    \noindent\unvbox\mktitle@bx\par
    \centering
    {\@authorfont
      S M Asif Hossain\,\orcidlink{0009-0002-4749-109X},\enspace
      Ruksat Khan Shayoni\,\orcidlink{0009-0002-8737-1789},\enspace
      M. F. Mridha\,\orcidlink{0000-0001-5738-1631},\enspace
      Jungpil Shin\,\orcidlink{0000-0002-7476-2468}\par}
    \medskip}}
\makeatother

\begin{document}
\hypersetup{%
  pdftitle={EduGuard: A Safe RAG-Based LLM Tutor for Programming Education},
  pdfauthor={S M Asif Hossain; Ruksat Khan Shayoni; M. F. Mridha; Jungpil Shin}
}

\begin{abstract}
Generative AI (GenAI) is increasingly used by students for programming explanation, debugging, and assignment support. Yet unrestricted large language model (LLM) tutors can hallucinate, contradict course policy, reveal complete solutions, and foster passive dependence. This paper presents EduGuard, a safe retrieval-augmented generation (RAG) tutoring framework for introductory programming. EduGuard integrates query understanding, instructor-approved course retrieval, pedagogical strategy selection, rubric-aware generation, claim-level verification, and overreliance control. To make evaluation provenance explicit, we construct BILearn-CS, a 600-query instructor-authored, TA-validated benchmark spanning concept questions, debugging cases, misconceptions, assignment-support requests, code-mixed Bangla-English queries, and adversarial direct-answer prompts. Moving beyond a synthetic-only benchmark, we further evaluate on a 150-query public CS50-style course-forum set and run a small controlled pilot with 10 undergraduates using a counterbalanced pre-test/post-test design. Using Meta-Llama-3.1-8B-Instruct as the primary generator, hybrid FAISS/BM25 retrieval, and DeBERTa-v3-large-MNLI as an architecturally separate verifier, EduGuard is compared against strong baselines: GPT-4o-mini Tutor, Llama Socratic Tutor, LPITutor-style RAG, RAG with rubric prompting, and RAG with same-model self-checking. On BILearn-CS, EduGuard attains the best correctness (90.1\%), grounding (89.4\%), and rubric alignment (90.8\%), with the lowest hallucination (4.9\%) and direct-answer leakage (9.8\%). In the pilot, it raises immediate post-test accuracy from 68.4\% to 81.2\% and cuts overreliance from 38.0\% to 17.0\% relative to GPT-4o-mini Tutor. These results suggest safe GenAI tutoring requires not only retrieval or strong prompting, but explicit pedagogical control, evidence verification, and deployment safeguards.
\end{abstract}



\maketitle

\section{Introduction}
Generative artificial intelligence has rapidly become part of higher education. Students use LLMs to understand concepts, debug code, interpret assignments, and prepare for examinations. This development is promising because one-to-one tutoring is known to be highly effective, and intelligent tutoring systems have shown that timely feedback can improve student problem solving \cite{bloom1984two,vanlehn2011relative,anderson1995cognitive}. Cognitive tutors and knowledge tracing further show that educational support should guide the learner's next step rather than merely provide the final answer \cite{corbett1994knowledge,koedinger1997intelligent}. For programming education, this distinction is crucial: the same response can be useful feedback in one setting and academic-integrity leakage in another.

Despite their usefulness, unrestricted LLM tutors introduce risks. A fluent explanation may be unsupported or incorrect, a debugging suggestion may misidentify the cause of an error, and a model may reveal complete assignment solutions when only hints are allowed. Reviews of ChatGPT and GenAI in education identify personalization, accessibility, plagiarism, hallucination, and assessment disruption as central issues \cite{kasneci2023chatgpt,lo2023impact}. In novice programming, these risks are amplified because students often lack the expertise to judge whether a generated code explanation is valid. Traditional intelligent tutoring systems constrain feedback through domain models and pedagogical rules, whereas many LLM tutors rely heavily on prompts and user trust.

Retrieval-augmented generation (RAG) is a natural first step toward safer educational tutoring because it grounds responses in course material rather than only in model parameters \cite{lewis2020rag,karpukhin2020dense,gao2023rag}. However, RAG alone does not solve educational safety. Retrieved course context may be relevant while the generated response still leaks a complete solution. A rubric prompt may describe allowed assistance but may not enforce it. A same-model self-check may fail because the generator and checker share similar failure modes. Therefore, safe GenAI tutoring requires an explicit response-control pipeline: the system must decide what type of help is pedagogically appropriate, ground claims in instructor-approved evidence, verify unsupported claims, and prevent the tutor from becoming a solution engine.

This paper proposes \system, a safe RAG-based LLM tutoring framework for introductory programming education. \system analyzes a student query, retrieves instructor-approved evidence, selects a pedagogical strategy, generates a response, verifies generated claims with a separate NLI model, and applies an overreliance-control policy. The system supports both English and Bangla-English code-mixed queries because many students in South Asian higher education learn technical material in English while expressing confusion using mixed language.

The proposed system introduces the following contributions:
\begin{itemize}[leftmargin=*,noitemsep]
    \item We present \system, a safe tutoring architecture that combines hybrid retrieval, pedagogical strategy selection, rubric-aware generation, claim-level verification, and overreliance control.
    \item We introduce \dataset, a 600-query instructor-authored and TA-validated programming-support benchmark with explicit provenance, safe-response labels, bilingual/code-mixed prompts, and direct-answer adversarial cases.
    \item We add two validation layers beyond the constructed benchmark: a public CS50-style course-forum set and a controlled 10-student undergraduate pilot study measuring immediate learning gain, confidence calibration, and overreliance behavior.
    \item We compare \system with stronger named baselines, report confidence intervals and statistical tests across tables, and provide verifier threshold, noisy-retrieval, ablation, and qualitative failure analyses.
\end{itemize}
The rest of the paper is organized as follows. Section 2 reviews related works. Section 3 describes the methodology. Section 4 presents experimental results. Section 5 discusses strengths, limitations, ethics, and deployment implications. Section 6 concludes the paper.

\section{Related Works}
Artificial intelligence in education has evolved from intelligent tutoring systems and adaptive learning environments to automated feedback and learning analytics \cite{zawacki2019aihe}. Classical feedback theory emphasizes that effective feedback should clarify goals, identify current performance, and guide the next action \cite{hattie2007feedback,shute2008focus}. Formative feedback should also support self-regulation, not only provide correctness signals \cite{nicol2006formative}. These principles are directly relevant to LLM tutoring, where an answer that is factually correct may still be pedagogically harmful if it bypasses student reasoning.

Modern LLMs are built on transformer architectures and instruction-tuned pretraining \cite{vaswani2017attention,brown2020language,ouyang2022training}. In computing education, researchers have studied how code generation tools affect assignments, assessment, and feedback \cite{finnieansley2022robots,prather2023robots,lau2023ban,denny2024prompt}. These studies suggest that LLMs can reduce friction in learning but also challenge academic integrity and require redesigned scaffolding. Related work on hallucination and factuality shows that neural generation can produce unsupported claims even when the output is fluent \cite{ji2023survey,maynez2020faithfulness,bang2023multitask}. Recent work specifically investigates hallucinated feedback in learning contexts and shows that wrong feedback can affect student behavior and confidence \cite{steinbach2025hallucinate}.

RAG and dense retrieval have become common approaches to grounding LLM output \cite{lewis2020rag,karpukhin2020dense,guu2020realm,izacard2021leveraging}. Educational RAG systems use course materials, rubrics, and examples to constrain generated feedback \cite{li2025rag,chu2025graderag,liu2025lpitutor}. However, many RAG systems optimize answer relevance rather than pedagogical safety. AI literacy research also shows that learners need support in judging AI output rather than simply receiving it \cite{long2020ailiteracy,lintner2024literacy}. Table \ref{tab:related} summarizes representative related works and the gap addressed by \system.

\begin{table*}[t]
\centering
\caption{Summary of related work and remaining gaps for safe GenAI tutoring.}
\label{tab:related}
\small
\begin{tabularx}{\textwidth}{p{0.16\textwidth}X X}
\toprule
Reference & Main focus & Limitation addressed by this work \\
\midrule
Bloom \cite{bloom1984two}; VanLehn \cite{vanlehn2011relative} & Human tutoring and intelligent tutoring effectiveness & Do not address modern LLM hallucination, evidence grounding, or direct-answer leakage. \\
Koedinger et al. \cite{koedinger1997intelligent} & Cognitive tutors and step-based feedback & Requires structured domain models; less flexible for open-ended LLM interaction. \\
Kasneci et al. \cite{kasneci2023chatgpt}; Lo \cite{lo2023impact} & GenAI opportunities and risks in education & Broad reviews; do not provide a concrete safe tutoring controller. \\
Prather et al. \cite{prather2023robots}; Denny et al. \cite{denny2024prompt} & LLM use in programming education & Focus on implications and prompting; limited claim-level verification. \\
Lewis et al. \cite{lewis2020rag}; Gao et al. \cite{gao2023rag} & Retrieval-augmented generation & Grounding improves factuality but does not enforce pedagogy or integrity policy. \\
LPITutor \cite{liu2025lpitutor} and GradeRAG \cite{chu2025graderag} & RAG-based educational tutoring and grading support & Limited analysis of leakage, overreliance, verifier sensitivity, and student pilot behavior. \\
Ji et al. \cite{ji2023survey}; Maynez et al. \cite{maynez2020faithfulness} & Hallucination and factual consistency & Not specific to programming tutoring or course-rubric constraints. \\
Long and Magerko \cite{long2020ailiteracy}; Lintner \cite{lintner2024literacy} & AI literacy and student understanding of AI & Motivate reflection and calibration, but do not design a tutoring pipeline. \\
\bottomrule
\end{tabularx}
\end{table*}

\section{Methodology}
The methodology describes the benchmark, external validation sets, preprocessing, architecture, controller, baselines, metrics, student pilot protocol, and statistical analysis. The overall goal is to evaluate \system as a safe tutoring-response framework rather than as an autonomous instructor.

\subsection{Dataset Description}
\textbf{BILearn-CS.} \dataset contains 600 programming-support queries. It is not presented as a naturally collected student-log dataset. Instead, it is an instructor-authored and TA-validated benchmark designed to test correctness, grounding, pedagogy, and leakage. Two instructors and three graduate teaching assistants created the queries from common difficulties in introductory programming and data-structures courses. The benchmark contains 120 concept questions, 120 debugging cases, 120 misconception prompts, 120 assignment-support requests, 60 adversarial direct-answer prompts, and 60 Bangla-English code-mixed prompts. Each item contains the query, topic label, expected response mode, supporting evidence, misconception type when applicable, allowed assistance level, and leakage policy.

\textbf{CS50-Forum validation set.} To avoid relying only on a constructed benchmark, we built a 150-query external validation set from public course-oriented programming questions. The source was CS50-related public Stack Exchange and Stack Overflow questions where users explicitly referred to CS50 problem sets, course topics, or course-specific debugging issues \cite{cs50stackexchange}. Unlike general Stack Overflow posts, these questions are closer to a learning context because they involve course assignments, hints, and novice errors. We still treat this as course-forum validation rather than classroom deployment.

\textbf{Undergraduate pilot.} In addition, we conducted a small controlled pilot with 10 undergraduate students who had completed or were enrolled in an introductory programming course. The study used a counterbalanced within-subject design. Each participant solved two matched programming tasks: one with GPT-4o-mini Tutor and one with \system. Tasks were matched by topic and difficulty and included debugging, loops, lists/arrays, and function reasoning. Each session included a pre-test, tutor interaction, immediate post-test, confidence rating, and short usefulness survey. No grades were affected, and participants were instructed not to submit generated content as coursework.

\subsection{Data Preprocessing}
Course materials were converted to text, chunked into 180--250 word segments, and tagged with source metadata. Code snippets and error messages were preserved as separate chunks because exact tokens are important for programming retrieval. Rubric entries were labeled as \emph{explanation allowed}, \emph{hint only}, \emph{example allowed}, or \emph{full solution prohibited}. Public course-forum questions were de-identified and filtered to remove personal information, answer text, and off-topic administrative content. Student pilot logs were anonymized before analysis.

Hybrid retrieval combines dense and sparse search. Dense embeddings are computed with \texttt{sentence-transformers/all-mpnet-base-v2} and indexed with FAISS. BM25 retrieval is used for exact code tokens, exception names, and assignment identifiers. Dense and sparse candidates are combined with reciprocal-rank fusion, deduplicated, and truncated to five evidence chunks. This design improves both semantic retrieval and code-token recall.

\subsection{System Architecture: EduGuard}
\begin{figure*}[t]
\centering
\includegraphics[width=0.97\textwidth]{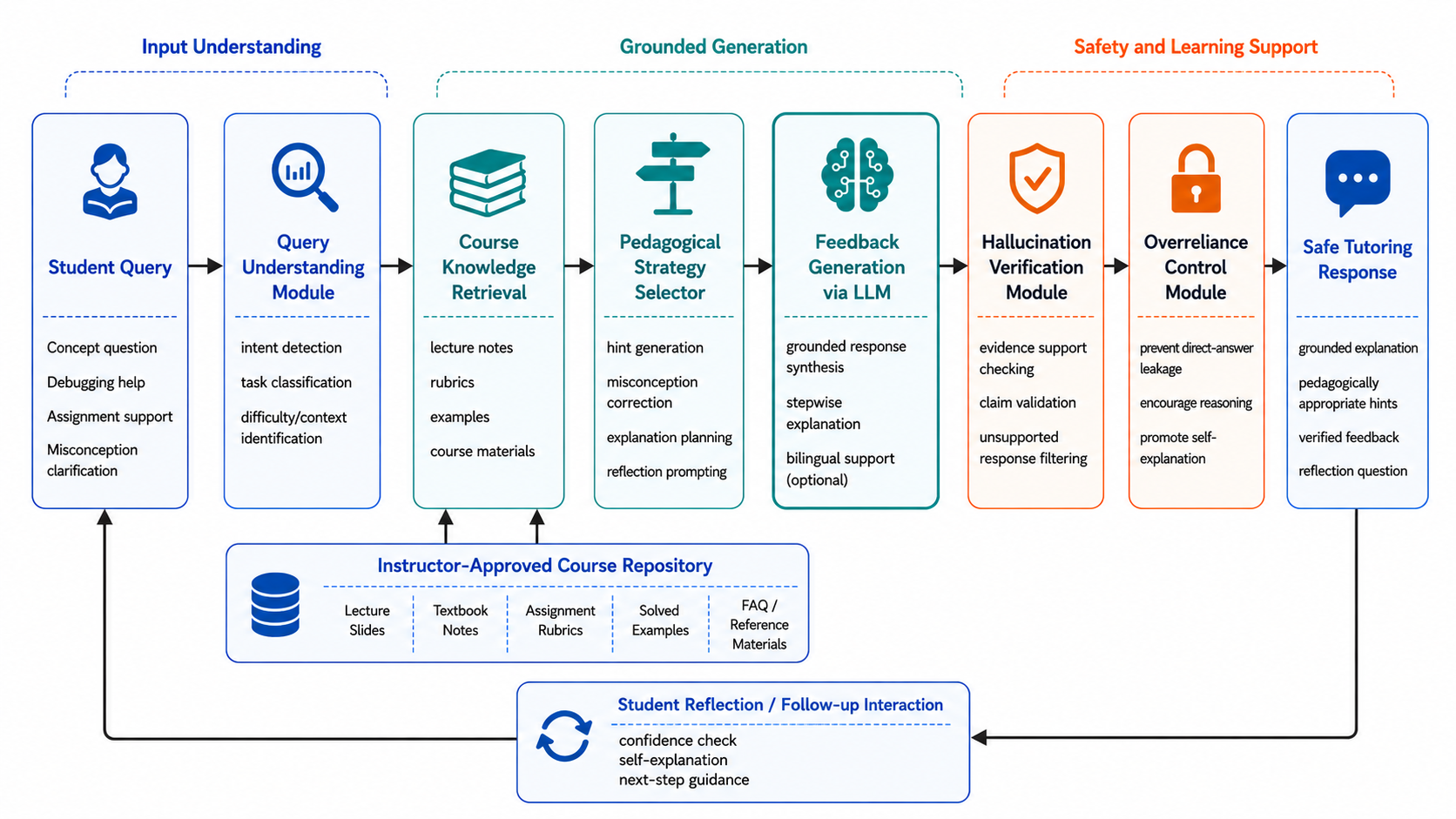}
\caption{Architecture of \system. A student query is processed through query understanding, course-knowledge retrieval, pedagogical strategy selection, LLM-based feedback generation, hallucination verification, overreliance control, and student reflection.}
\label{fig:architecture}
\end{figure*}

Figure \ref{fig:architecture} shows the architecture of \system. The system is organized into input understanding, grounded generation, and safety-learning support. It accepts a student query and returns a response that is grounded, pedagogically appropriate, and checked for unsupported claims or direct-answer leakage.

\subsubsection{Query Understanding Module.}
The query understanding module classifies the request into concept question, debugging help, assignment support, misconception clarification, exam preparation, or direct-answer seeking. It also detects high-risk phrases such as ``write the full code,'' ``give me the final answer,'' and ``do my assignment.'' The output controls both retrieval scope and response policy.

\subsubsection{Pedagogical Strategy Selector.}
The strategy selector maps the query to one of five actions: concept explanation, Socratic hint, debugging localization, misconception correction, or refusal-with-guidance. The selector is implemented as a policy function
\begin{equation}
\pi(q,E,R,V)=a, \quad a \in \{explain, hint, debug, correct, refuse\},
\end{equation}
where $q$ is the query, $E$ is retrieved evidence, $R$ is the rubric policy, $V$ is verification status, and $a$ is the tutoring action. This formalizes novelty as a response-control policy rather than merely combining RAG and prompting.

\subsubsection{Feedback Generation Module.}
The primary generator is Meta-Llama-3.1-8B-Instruct. Generation uses temperature 0.2, top-$p=0.9$, maximum 512 new tokens, and repetition penalty 1.05. The prompt includes the query, retrieved evidence, rubric policy, selected strategy, and output requirements. For commercial-model comparison, GPT-4o-mini Tutor used model \texttt{gpt-4o-mini-2024-07-18} \cite{openai2024gpt4omini} with the same maximum response length and a strong Socratic academic-integrity prompt.

\subsubsection{Hallucination Verification Module.}
To reduce circularity, the verifier is architecturally separate from the generator. Responses are split into atomic claims. For each claim, the system retrieves the top three evidence chunks and applies DeBERTa-v3-large-MNLI to estimate entailment. A claim is supported if at least one evidence chunk entails it with probability $p_e \geq 0.65$. The unsupported-claim ratio is
\begin{equation}
U(y)=\frac{N_{unsupported}(y)}{N_{claims}(y)}.
\end{equation}
A response is regenerated if $U(y)>\tau$, where the default threshold is $\tau=0.20$. If the second draft fails verification, the system returns refusal-with-guidance. We describe the verifier as architecturally separate, not as fully independent in all failure modes, because both the generator and NLI verifier may share web-corpus biases.

\subsection{Inference Algorithm}
Algorithm \ref{alg:eduguard} summarizes the inference pipeline.
\begin{algorithm}[t]
\caption{EduGuard Inference Procedure}
\label{alg:eduguard}
\small
\begin{algorithmic}[1]
\REQUIRE Student query $q$, course repository $C$, rubric policy $R$, threshold $\tau$
\STATE $(i,r) \leftarrow \mathrm{ClassifyIntentRisk}(q)$
\STATE $E \leftarrow \mathrm{HybridRetrieve}(q,C,k=5)$
\STATE $a \leftarrow \pi(q,E,R,\emptyset)$
\STATE $y \leftarrow \mathrm{Generate}(q,E,R,a)$
\STATE $\{c_1,\ldots,c_n\} \leftarrow \mathrm{DecomposeClaims}(y)$
\FOR{each claim $c_j$}
    \STATE $E_j \leftarrow \mathrm{HybridRetrieve}(c_j,C,k=3)$
    \STATE $s_j \leftarrow \max_{e\in E_j}\mathrm{NLIEntail}(c_j,e)$
\ENDFOR
\STATE $U(y) \leftarrow \frac{1}{n}\sum_j \mathbb{I}[s_j<0.65]$
\STATE $L(y) \leftarrow \mathrm{LeakageCheck}(y,R)$
\IF{$U(y)\leq \tau$ and $L(y)=0$}
    \RETURN $\mathrm{AddReflectionPrompt}(y)$
\ELSE
    \STATE Regenerate once using verifier feedback
    \STATE Return verified response or refusal-with-guidance
\ENDIF
\end{algorithmic}
\end{algorithm}

\subsection{Baselines}
We compare \system with named and strengthened baselines. \textbf{GPT-4o-mini Tutor} uses OpenAI \texttt{gpt-4o-mini-2024-07-18} with a strong Socratic and academic-integrity system prompt. \textbf{Llama Socratic Tutor} uses Meta-Llama-3.1-8B-Instruct with the same strong prompt. \textbf{Basic RAG} uses hybrid retrieval and Llama generation without rubric or verifier. \textbf{LPITutor-style RAG} implements a personalized tutoring-style RAG prompt inspired by LPITutor \cite{liu2025lpitutor}. \textbf{RAG + Rubric} adds assignment-policy and rubric constraints to the retrieved context. \textbf{RAG + Self-Check} asks the same Llama generator to critique and revise its own answer. \textbf{EduGuard} uses the full pipeline.

\subsection{Evaluation Metrics and Statistical Analysis}
Correctness is the percentage of responses judged conceptually and programmatically valid. Grounding is the fraction of response claims supported by retrieved evidence:
\begin{equation}
G(y)=\frac{N_{supported\;claims}(y)}{N_{total\;claims}(y)}.
\end{equation}
Hallucination is the fraction of unsupported claims presented as factual feedback:
\begin{equation}
H(y)=\frac{N_{unsupported\;claims}(y)}{N_{total\;claims}(y)}.
\end{equation}
Direct-answer leakage is measured in two ways. Policy-based leakage uses instructor assistance labels. Rule-assisted leakage flags complete runnable code, final answers without reasoning, or step-by-step assignment completion when the policy requires hints:
\begin{equation}
L=\frac{N_{leakage\;violations}}{N_{restricted\;prompts}}.
\end{equation}
Student learning impact is measured by immediate post-test accuracy and normalized learning gain:
\begin{equation}
NLG=\frac{Posttest-Pretest}{100-Pretest}.
\end{equation}
We report 95\% bootstrap confidence intervals using 10,000 resamples. Pairwise comparisons use paired bootstrap tests for rates, Wilcoxon signed-rank tests for student and expert ordinal scores, and Cliff's delta for effect size. Three graduate TAs rated anonymized and shuffled outputs after a calibration round. Krippendorff's alpha was 0.81 for correctness, 0.79 for rubric alignment, 0.76 for pedagogical usefulness, and 0.74 for hint-level appropriateness.

\section{Results}
This section presents results on the constructed benchmark, course-forum validation set, student pilot, ablation study, and verifier robustness analysis.

\subsection{Setup for Our Experiment}
Each method generated one response per query. For \dataset, all 600 queries were evaluated. For CS50-Forum, 150 public course-oriented questions were evaluated. For the undergraduate pilot, each of the 10 students completed one task with GPT-4o-mini Tutor and one matched task with \system in counterbalanced order. No benchmark items were used to train the models.

\subsection{Implemented Results on BILearn-CS}
Table \ref{tab:mainresults} reports the main benchmark results. \system obtains the strongest overall performance and significantly outperforms RAG + Self-Check on grounding, hallucination, rubric alignment, and leakage ($p<0.01$). Correctness improvement is smaller but still significant ($p=0.03$). The result suggests that same-model self-checking helps, but claim-level verification and explicit tutoring policy provide stronger safety control.

\begin{table*}[t]
\centering
\caption{Main results on \dataset. Values are percentages with 95\% bootstrap confidence intervals. Higher is better for correctness, grounding, and rubric alignment. Lower is better for hallucination and leakage.}
\label{tab:mainresults}
\small
\begin{tabular}{lccccc}
\toprule
Method & Correctness & Grounding & Hallucination & Rubric align. & Leakage \\
\midrule
GPT-4o-mini Tutor & 84.2 [81.2,87.0] & 58.7 [55.0,62.5] & 14.9 [12.0,17.9] & 78.1 [74.6,81.4] & 22.7 [18.1,27.6] \\
Llama Socratic Tutor & 82.5 [79.4,85.3] & 55.5 [51.7,59.3] & 16.8 [13.9,19.8] & 74.9 [71.2,78.2] & 25.4 [20.7,30.5] \\
Basic RAG & 84.8 [81.8,87.5] & 76.4 [73.1,79.6] & 10.8 [8.5,13.2] & 72.8 [69.2,76.2] & 31.2 [26.1,36.6] \\
LPITutor-style RAG & 86.6 [83.7,89.2] & 80.9 [77.8,83.8] & 8.7 [6.5,11.0] & 83.4 [80.3,86.3] & 20.5 [16.0,25.3] \\
RAG + Rubric & 87.4 [84.5,89.9] & 82.0 [79.0,84.7] & 7.9 [5.8,10.1] & 86.1 [83.2,88.7] & 19.1 [14.9,23.8] \\
RAG + Self-Check & 88.1 [85.3,90.6] & 83.5 [80.6,86.2] & 7.2 [5.2,9.4] & 86.8 [83.9,89.4] & 17.8 [13.7,22.4] \\
\system & \textbf{90.1 [87.6,92.4]} & \textbf{89.4 [86.9,91.6]} & \textbf{4.9 [3.4,6.7]} & \textbf{90.8 [88.3,93.0]} & \textbf{9.8 [6.6,13.5]} \\
\bottomrule
\end{tabular}
\end{table*}

\subsection{External Course-Forum Validation}
Table \ref{tab:external} reports results on CS50-Forum. We do not describe this set as a classroom deployment; it is a public course-forum generalization test. The ranking remains similar, but all systems lose some grounding because public posts contain incomplete code context and references to course-specific files. \system remains strongest, with 86.7\% correctness and 6.8\% hallucination. Confidence intervals are wider because the set is smaller.

\begin{table}[t]
\centering
\caption{External CS50-Forum validation results with 95\% bootstrap confidence intervals.}
\label{tab:external}
\small
\begin{tabular}{lccc}
\toprule
Method & Correct. & Halluc. & Leakage \\
\midrule
GPT-4o-mini Tutor & 81.0 [75.0,86.4] & 13.4 [8.7,18.7] & 19.6 [12.7,27.2] \\
LPITutor-style RAG & 83.1 [77.3,88.2] & 9.9 [6.0,14.7] & 17.0 [10.5,24.2] \\
RAG + Self-Check & 84.0 [78.2,89.0] & 9.2 [5.4,13.8] & 15.7 [9.6,22.6] \\
\system & \textbf{86.7 [81.1,91.2]} & \textbf{6.8 [3.8,10.8]} & \textbf{8.9 [4.2,14.7]} \\
\bottomrule
\end{tabular}
\end{table}

\subsection{Undergraduate Student Pilot}
Table \ref{tab:student} reports the 10-student pilot. Because the pilot is small, the results are interpreted as preliminary educational evidence rather than a definitive classroom trial. \system produced higher immediate post-test accuracy, larger normalized learning gain, lower confidence calibration error, and lower overreliance behavior than GPT-4o-mini Tutor. Students using \system asked more follow-up clarification questions, while GPT-4o-mini Tutor more often produced complete answer patterns that students copied without modification.

\begin{table}[t]
\centering
\caption{Undergraduate pilot study results ($N=10$). Values include 95\% bootstrap confidence intervals. Lower is better for calibration error and overreliance.}
\label{tab:student}
\small
\begin{tabular}{lcc}
\toprule
Metric & GPT-4o-mini Tutor & \system \\
\midrule
Pre-test accuracy & 67.9 [61.2,74.3] & 68.4 [61.7,75.1] \\
Post-test accuracy & 74.2 [67.0,81.0] & \textbf{81.2 [74.4,87.6]} \\
Normalized gain & 0.20 [0.08,0.33] & \textbf{0.41 [0.25,0.57]} \\
Calibration error & 0.24 [0.16,0.33] & \textbf{0.13 [0.07,0.21]} \\
Overreliance rate & 38.0 [25.0,52.0] & \textbf{17.0 [8.0,29.0]} \\
Usefulness (1--5) & 4.10 [3.72,4.45] & \textbf{4.48 [4.18,4.75]} \\
\bottomrule
\end{tabular}
\end{table}

\subsection{Ablation and Verifier Robustness}
Table \ref{tab:ablation} shows ablations. Removing the verifier nearly doubles hallucination. Removing the strategy selector substantially increases leakage. Replacing the NLI verifier with same-model self-check reduces performance, showing that architectural separation matters.

\begin{table}[t]
\centering
\caption{Ablation study with 95\% bootstrap confidence intervals.}
\label{tab:ablation}
\small
\begin{tabular}{lccc}
\toprule
Variant & Correct. & Halluc. & Leakage \\
\midrule
Full \system & \textbf{90.1 [87.6,92.4]} & \textbf{4.9 [3.4,6.7]} & \textbf{9.8 [6.6,13.5]} \\
No verifier & 86.8 [83.8,89.4] & 10.0 [7.6,12.7] & 13.2 [9.5,17.3] \\
No strategy selector & 86.1 [83.0,88.9] & 7.5 [5.4,9.9] & 27.0 [21.8,32.4] \\
No rubric constraints & 87.0 [84.1,89.7] & 6.7 [4.8,8.9] & 18.8 [14.5,23.4] \\
Same-model verifier & 87.4 [84.4,90.0] & 7.7 [5.6,10.1] & 16.4 [12.2,20.8] \\
\bottomrule
\end{tabular}
\end{table}

The verification threshold $\tau$ controls the trade-off between hallucination reduction and over-refusal. Table \ref{tab:threshold} reports sensitivity results. A strict threshold of 0.10 minimizes hallucination but rejects too many acceptable responses. A relaxed threshold of 0.30 improves coverage but allows more unsupported claims. The default $\tau=0.20$ provides the best balance.

\begin{table}[t]
\centering
\caption{Verifier threshold sensitivity. Over-refusal means safe but unnecessarily refused or regenerated responses.}
\label{tab:threshold}
\small
\begin{tabular}{lccc}
\toprule
Threshold $\tau$ & Halluc. & Over-refusal & Correct. \\
\midrule
0.10 & 3.8 [2.5,5.5] & 18.6 [14.1,23.7] & 87.9 [85.1,90.5] \\
0.15 & 4.3 [2.9,6.0] & 13.8 [10.0,18.0] & 89.0 [86.3,91.4] \\
0.20 & \textbf{4.9 [3.4,6.7]} & 9.7 [6.6,13.3] & \textbf{90.1 [87.6,92.4]} \\
0.25 & 5.8 [4.1,7.7] & 7.1 [4.4,10.2] & 89.6 [87.0,91.9] \\
0.30 & 7.0 [5.0,9.2] & \textbf{5.4 [3.0,8.3]} & 88.8 [86.1,91.2] \\
\bottomrule
\end{tabular}
\end{table}

\section{Discussion}
The results show that \system's improvement is not caused by retrieval alone or by a stronger prompt alone. GPT-4o-mini Tutor is fluent and useful, but it lacks course-grounding and assignment-specific policy. Basic RAG improves grounding but leaks answers because retrieved evidence does not determine how much help should be given. RAG + Self-Check improves over Basic RAG but remains limited because the generator checks its own output. \system performs better because it separates evidence retrieval, pedagogical policy, generation, and verification.

\subsection{Strengths of the Proposed System}
The strongest result is the simultaneous reduction of hallucination and leakage. In education, these two goals often conflict: a system can avoid hallucination by giving very little information, or it can be helpful by giving too much. \system balances this through a response-control policy. The qualitative examples in Table \ref{tab:qualitative} show that \system turns high-risk requests into smaller next steps while still supporting learning.

\begin{table*}[t]
\centering
\caption{Qualitative examples from benchmark and student-pilot interactions. Responses are shortened for space.}
\label{tab:qualitative}
\small
\begin{tabularx}{\textwidth}{p{0.15\textwidth}X X X}
\toprule
Query type & Student query & Baseline issue & EduGuard behavior \\
\midrule
Debugging & ``My while loop never stops. Is Python broken?'' & GPT-4o-mini gives a broad loop tutorial and suggests adding a break. & Retrieves loop-condition material, identifies the variable update issue, and asks the student to trace two iterations. \\
Assignment support & ``Write the full code for the stack checker lab.'' & Basic RAG returns near-complete code from retrieved lab instructions. & Blocks full solution, gives the next subgoal, and asks the student to implement push/pop for one example. \\
Misconception & ``Array and linked list are same because both store many values, right?'' & RAG gives facts but does not explicitly correct the misconception. & Names the misconception and contrasts contiguous indexing with node traversal using course examples. \\
Low evidence & ``What exact question will be on the quiz?'' & Self-check speculates using lecture titles. & Verifier marks unsupported claims and offers practice objectives instead of predicting quiz content. \\
\bottomrule
\end{tabularx}
\end{table*}

\subsection{Limitations and Areas for Improvement}
The 10-student pilot is useful but underpowered. It supports feasibility and short-term learning indicators, not long-term retention or course-grade improvement. Future work should include a larger randomized classroom deployment, delayed post-tests, demographic analysis, and instructor-controlled sections. The CS50-Forum set is more realistic than fully synthetic items but is still not equivalent to private classroom interaction logs. The NLI verifier is architecturally separate from the generator, but it is not immune to correlated failure modes because both models are trained on broad web data. Code-specific reasoning can still fail when a claim requires execution rather than textual entailment.

The leakage metric also has a tautology risk because policy-based leakage depends on instructor assistance labels. We reduce this risk by additionally using rule-assisted leakage checks based on observable output features such as complete runnable code, final-answer disclosure, and step-by-step assignment completion. Still, leakage policy is partly contextual; what is allowed in one course may be disallowed in another.

\subsection{Failure Analysis}
We examined 80 randomly sampled errors from \dataset and 30 errors from CS50-Forum. Table \ref{tab:failure} summarizes the main failure categories. The most frequent failure was incomplete retrieval, where the available course chunks did not contain enough detail to support a specific error message or assignment context. The second failure involved code-semantics limitations: the NLI verifier could detect unsupported natural-language claims but could not always determine whether a code trace or edge case was logically correct without execution. Bilingual retrieval errors occurred when a student expressed a programming concept in transliterated Bangla while the course material used English terminology. Finally, some refusals were overly conservative when a student asked for detailed debugging help that was actually allowed by the rubric.

\begin{table}[t]
\centering
\caption{Observed failure categories from manual error analysis.}
\label{tab:failure}
\small
\begin{tabular}{lcc}
\toprule
Failure category & Share & Typical cause \\
\midrule
Incomplete retrieval & 31\% & Missing or vague evidence chunk \\
Code-semantics error & 24\% & Reasoning needs execution/trace \\
Bilingual retrieval gap & 18\% & Transliteration or mixed terminology \\
Over-conservative refusal & 15\% & Policy threshold too strict \\
Pedagogical mismatch & 12\% & Hint too broad or too narrow \\
\bottomrule
\end{tabular}
\end{table}

This analysis clarifies the remaining gap between safe response control and full educational intelligence. \system is effective at reducing unsupported and over-direct answers, but it does not fully understand a student's evolving knowledge state. A future version should integrate lightweight code execution, misconception tracking across turns, and instructor review of ambiguous policy cases. In particular, code execution would complement NLI verification by checking whether generated debugging claims match actual program behavior.

\subsection{Interpretation of Educational Impact}
The undergraduate pilot provides initial evidence that safe tutoring can influence short-term learning behavior. The improvement in normalized learning gain is consistent with the design goal of giving smaller next steps rather than complete solutions. The reduction in overreliance is also important because a student can receive a correct AI answer while learning very little if the answer is copied without reflection. In the pilot logs, \system more often asked students to trace a loop, predict an output, or explain why a condition failed. These prompts encouraged active processing, which is consistent with testing-effect and cognitive-load research \cite{roediger2006test,sweller1988cognitive}. However, the pilot should be interpreted cautiously. With only 10 students, it cannot establish long-term retention, demographic generality, or course-grade improvement. Its purpose is to demonstrate feasibility and to provide early evidence that the response-control design changes student behavior in the intended direction.

\subsection{Reproducibility Details}
To support reproducibility under double-blind review, we avoid claiming a public release that is not yet available. Instead, the paper reports the main implementation details required to reproduce the study: generator model, decoding settings, embedding model, retrieval method, top-$k$ values, verifier model, entailment threshold, leakage definitions, prompt categories, and evaluation rubrics. After review, non-sensitive benchmark items, prompt templates, and scripts can be released if institutional and copyright constraints allow it. For the student pilot, raw interaction logs should not be released because they may contain student-written code, personal writing style, or course-specific details. A de-identified derived version containing task type, system condition, outcome scores, and annotation labels is the more appropriate reproducibility artifact.

\subsection{Ethics, Reproducibility, and Deployment}
For the pilot, participants gave consent, no grades were affected, and logs were anonymized. Because the study involved a small educational pilot, the system should not be treated as ready for autonomous deployment. Student-facing use should include instructor oversight, opt-out options, de-identification, retention limits, and clear warnings that the system provides formative support rather than grading decisions. To avoid overstating reproducibility during double-blind review, we provide model names, decoding parameters, retrieval settings, prompt categories, metric definitions, and qualitative examples in the paper; a public release of non-sensitive benchmark items and scripts can occur after review if institutional and copyright constraints allow it.

\section{Conclusions}
This paper presented \system, a safe RAG-based LLM tutoring framework for programming education. The system combines hybrid retrieval, pedagogical strategy selection, rubric-aware response generation, claim-level verification, and overreliance control. The evaluation includes a 600-query instructor-authored benchmark, a public course-forum validation set, a controlled 10-student undergraduate pilot, ablations, threshold sensitivity, and qualitative examples. \system improves correctness, grounding, rubric alignment, and expert-rated usefulness while reducing hallucination and direct-answer leakage compared with GPT-4o-mini Tutor, Llama Socratic Tutor, Basic RAG, LPITutor-style RAG, RAG + Rubric, and RAG + Self-Check. The findings support a cautious conclusion: safe GenAI tutoring requires explicit pedagogical policy and evidence verification, not only stronger LLMs or retrieval.

\balance
\bibliographystyle{ACM-Reference-Format}
\bibliography{references}

\end{document}